\documentclass[a4paper]{emsart}
\usepackage{emsartredef}
\usepackage{graphicx}%
\usepackage{dcolumn}%
\usepackage{bm}%

\usepackage[utf8]{inputenc}
\usepackage[T1]{fontenc}
\usepackage{mathptmx}
\usepackage{etoolbox}
\usepackage{hyperref}
\usepackage{subfig}
\usepackage[capitalise,nameinlink]{cleveref}
\crefname{equation}{}{}
\makeatletter
\let\@fnsymbol@latex\@fnsymbol
\let\@fnsymbol\@alph
\makeatother
	\volumeyear{2024}
	\journalname{Electromagnetic Science}
	\DOI{10.23919/emsci.2024.0009}
	\journaltype{Original Article}
	\title[On the Path to High-temperature Josephson Multi-junction Devices]{On the Path to High-temperature Josephson Multi-junction Devices}
	\author{%
			Xu~Wang\affilnums{1}, 
			Fucong~Chen\affilnums{1},
			Zefeng~Lin\affilnums{1},
 			Changhong~Yuan\affilnums{1},
 			Shibing~Tian\affilnums{1},
			Chunguang~Li\affilnums{1},
			Victor~Kornev\affilnums{2}, and 
			Nikolay~Kolotinskiy\affilnums{2}}
	\affiliation{%
		\affilnum{1}Institute of Physics, Chinese Academy of Sciences, Beijing, China\\
		\affilnum{2}Faculty of Physics, Lomonosov Moscow State University, Moscow, Russia
	}
	\corrauth{Xu Wang}
	\email{risingsunwx@iphy.ac.cn}
	\fundinginfo{This work was supported by National Key R\&D Program of China 2022YFA1603900 and in part by Russian Science Foundation (RSCF) grant no. 19-72-10016-P.}
	\receivetime{Received March 11, 2024}
	\accepttime{Accepted August 30, 2024}
	\authorcopy{The Authors}
	
	\abstract{
		We report our progress in the high-temperature
		superconductor (HTS) Josephson junction fabrication process founded on
		using a focused helium ion beam damaging technique and discuss the
		expected device performance attainable with the HTS multi-junction
		device technology. Both the achievable high value of characteristic
		voltage $V_c=I_cR_N$ of
		Josephson junctions and the ability to design a large number of
		arbitrary located Josephson junctions allow narrowing the existing gap
		in design abilities for LTS and HTS circuits even with using a single
		YBCO film layer. A one-layer topology of active electrically small
		antenna is suggested and its voltage response characteristics are
		considered.}
	
	\keywords{Josephson junctions, high-temperature
		superconductors, YBCO, focused helium ion beam fabrication technique,
		Josephson multi-junction devices, SQUID, active electrically small
		antenna}
		
\begin{document}	
		
	\maketitle
	
	\section{Introduction}
	
	Superconductive electronics technologies leverage the distinctive
	macroscopic quantum phenomena exhibited by superconductors. These
	phenomena include superconductivity, magnetic flux quantization, and
	the effect of macroscopic quantum interference and Josephson
	effect \cite{Cardwell2022,Likharev1986,Seidel2015}, which are applicable across a wide range of
	frequencies from direct current (dc) to very high frequencies. In fact, 
	degradation of the Josephson junction properties relevant 
	to their applications starts only at Josephson oscillation frequencies exceeding the characteristic Josephson junction 
	frequency
	\begin{equation}
	\label{eq:CharFreq}
	F_{c} = V_{c}/\Phi_{0} = I_{c}R_{N}/\Phi_{0},
	\end{equation}
	where $V_{c}$, $I_c$ and $R_N$
	are the junction characteristic voltage, critical current, and
	resistance in the normal state, respectively, and $\Phi_{0} = h/2e$
	represents the flux quantum. $F_{c}$ can reach hundreds of gigahertz
	and beyond. In addition, losses in superconductors remain low until
	reaching the gap frequency \cite{Seidel2015} (for example, the gap frequency of Nb
	is about $700\text{ GHz}$). All these properties enable the development of a variety of superconductor devices and circuits \cite{Cardwell2022,Seidel2015} with characteristics unattainable for semiconductor 
	electronic
	s technology
	. Among those, there is diverse range of digital and mixed-signal
	superconductive circuits, namely analog-to-digital converters (ADC)
	\cite{Mukhanov2004,Inamdar2009,Vernik2007,Sarwana2009,Gupta2011} and integrated digital circuits \cite{Herr2015}, including prototypes
	of microprocessors \cite{Yamanashi2007,Filippov2012,Ando2016} , which can operate at frequencies with
	tens of gigahertz and even higher \cite{Filippov2024} These devices are powered by
	the Rapid Single Flux Quantum (RSFQ) logic \cite{Mukhanov1987,Polonsky1993,Bunyk2001,Mukhanov2011,Kirichenko2011,Volkmann2012,Herr2011,Takeuchi2013,Holmes2013}, which operates
	using single flux quanta of magnetic flux $\Phi_{0} = h/2e$ for data
	signals and clock. In the RSFQ circuits, a single switching event of an
	overdamped Josephson junction (at
	$R_{N}C < 1/{\left( 2\pi F_{c} \right)}$, where $C$ is the
	capacitance of a junction \cite{Mukhanov2004}) with a typical
	$I_{c}\sim100\ \mu\text{A}$ dissipates an extremely low energy of
	$E_{J} = \Phi_{0}I_{c}\sim10^{-19}\text{ J}$. Moreover, the switching
	time $\tau$ of the junction is about $10^{- 12}\text{ s}$, allowing for
	clock frequencies in the hundreds of gigahertz range (\textit{e.g.} see
	\cite{Filippov2024}).
	
	Multiple Josephson junction devices can essentially improve the
	sensitivity and bandwidth of superconducting analog devices. Among them
	are SQUID-based array amplifiers \cite{Huber2001}, traveling wave parametric
	amplifiers \cite{Macklin2015,White2015,Bell2015,Zorin2016}, as well as Superconducting Quantum Filters (SQIFs) \cite{Caputo2005,Oppenlander2001} and 
	Superconducting Quantum Arrays (SQAs) \cite{Kornev2014} allowing implementation 
	of highly sensitive active electrically small
	antennas (ESAs) capable of providing both the reception and
	amplification of an incident electromagnetic wave \cite{Kornev2016,Kornev2017}. Such
	broadband ESAs can improve overall performance of superconductive
	broadband digital-rf receivers based on direct signal digitization
	allowing then digital extraction of customizable sub-bands \cite{Brock2001, Gupta2002}.
	As reported in \cite{Vernik2007a,Gupta2011a,Sarwana2011}, 
	the receivable signal frequency can be
	ranged from VHF to K frequency bands.
	
	These advances in the superconductive electronics have been achieved
	mainly by the use of low-temperature superconductors (LTS) and first of
	all a niobium fabrication process based on using tunnel Josephson
	junctions \cite{Seidel2015} with aluminum oxide barrier (\textit{e.g.} see \cite{SEEQC}). 
	The implementation of high-temperature superconductor (HTS) devices, other
	than the simplest ones containing only a few Josephson junctions, poses
	a stiff challenge because of substantial design restrictions within the
	grain-boundary junction fabrication techniques (with either bi-crystal
	or step-edge or ramp-edge junctions) \cite{Seidel2015} which have been generally
	used until now. In spite of that, much simpler array-type HTS structures
	but composed of great deal of Josephson junctions were nevertheless
	realized using fabrication technique of step-edge Josephson junctions
	\cite{Taylor2016}. Such a SQIF-like 2D network containing in whole 54000 SQUIDs
	showed a magnetic field to voltage transfer function with a central peak
	value of approximately $8\text{ mV}$ and a highly linear central peak portion
	around $4.5\text{ mV}$ \cite{Taylor2016}.
	
	Because of the short coherence length and small capacitance, ideal HTS SIS junction with hysteresis is very difficult to realize. Most HTS junctions exhibited a current-voltage (IV) curve similar to that of the LTS weak-link 
	junctions. Various types HTS Josephson junctions such as 
	bicrystal, step-edge, ramp-edge and ion-beam damaged 
	junctions have been realized and compared in reviews and 
	books \cite{Seidel2015}. Among these techniques, the ion-beam damaging technique can obtain a controllable local suppression
	of superconducting properties of HTS films down to metallic or even to dielectric properties. When a focused ion 
	beam is used, it enables forming a narrow barrier across a 
	superconducting strip to realize a weak link between the 
	strip parts and hence a Josephson junction with characteristics depending on irradiation dose \cite{Cho2015,Chen2022}. Such a technique provides basis for a highly promising fabrication 
	process enabling creation of high-quality multi-junction 
	HTS devices. In fact, one can refer to \cite{Ouanani2016,Pawlowski2018,Couedo2019} 
	where the ion-beam damaging impact on HTS films of
	YBa\textsubscript{2}Cu\textsubscript{3}O\textsubscript{7-x} (YBCO) has
	been already used to fabricate series SQIFs consisting of 300 to 2000 dc
	SQUIDs. Being in an unshielded magnetic environment, these devices were
	able to measure magnetic signals as low as a few pT in a frequency band
	ranged up to $1.125\text{ GHz}$. Besides, the devices showed a linear voltage
	response over 7 decades in RF power, that is a highly encouraging
	achievement.
	
	In this paper, we report our progress in the HTS Josephson junction
	fabrication process founded on using a focused helium ion beam damaging
	effect and discuss the expected device performance attainable for the
	HTS multi-junction device technology.
	
	\section{Progress in HTS Josephson junction fabrication}
	
	The localized impact of the helium ion irradiation on a
	YBa\textsubscript{2}Cu\textsubscript{3}O\textsubscript{7-x} (YBCO) HTS
	film can provide a local damage of the film structure resulting in local
	degradation of the film superconducting properties with irradiation dose
	down to normal metal conductivity and then to dielectric behavior
	\cite{Cho2015}. By using the focused helium ion beam, the damaged film area
	across a film strip can be made very narrow to provide weak electric
	coupling between the strip parts of metallic or tunnelling types that
	enables forming Josephson junctions which can have properties of SNS to
	SIS type junctions \cite{Mueller2019}.
	
	We used YBCO films deposited on STO substrates by pulsed laser
	deposition (PLD) technique. The films deposited at high oxygen pressure
	are grown epitaxially together with the substrate and therefore are
	highly crystalline with a highly uniform crystal orientation.
	Furthermore, the films deposited at high oxygen pressure also
	demonstrate a reasonably low microwave surface resistance of about $0.7\text{ m}\Omega$ per square ($77\text{ K}$, $9.4\text{ GHz}$) \cite{Xiong2023}.
	
	To fabricate and study the focused helium ion beam Josephson junctions,
	4 $\mu$m-wide film bridges (with 4 electrodes to enable four-point
	measurement of IV curve) were patterned with electron beam lithography
	(EBL) on a $35\text{ nm}$ thick YBCO thin film with an \textit{in situ} deposited Au layer
	on the top. For the EBL process, the acceleration voltage 
	of electrons was $10\text{ keV}$ and the beam current was about
	$0.2\text{ nA}$. The Au layer on the bridges was removed by wet etching. Next, a
	$35\text{ kV}$ helium beam was used to write
	on the bridges by crossing the $4$-micrometer-wide superconducting film bridges, thereby forming narrow damaged areas and hence 	to form Josephson junctions with irradiation dose changing from $100$ ions/nm to $700$ ions/nm. Resistance of the damaged area as a barrier
	between the film strip parts shows the temperature dependence
	corresponding to the metal type barrier at a lower dose and gradually
	varying to the insulator like barrier with the dose increase.

\begin{figure}[t!]
		\centering
		\subfloat[]{\includegraphics[width=7.2cm]{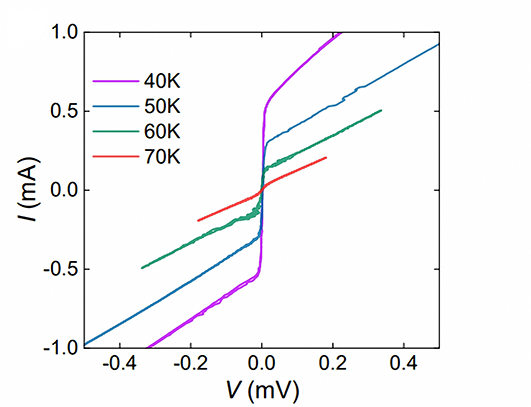}}\\
		\subfloat[\label{fig:HTS200Rn}]{\includegraphics[width=7.2cm]{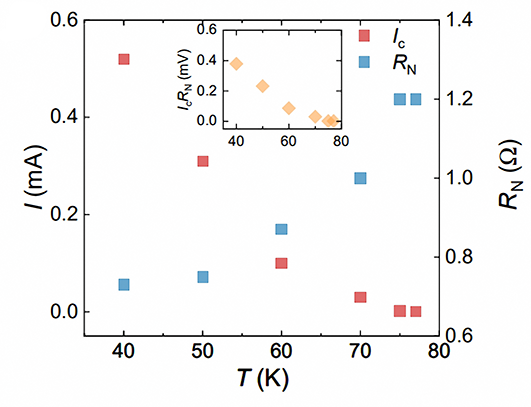}}
		\caption {\label{fig:HTS200Curves}IV curves (a), critical current $I_c$, normal
			resistance $R_N$ (b), and their product (characteristic voltage)
			$I_cR_N$ (inset) of the Josephson junctions fabricated using a helium dose of $200\text{ ions/nm}$, which are measured at different temperatures: $40$, $50$, $60$ and $70\text{ K}$.}
	\end{figure}

	\Cref{fig:HTS200Curves} shows IV curves (a), critical current $I_{c}$, normal
	resistance $R_{N}$, and their product (characteristic voltage)
	$I_{c}R_{N}$ (b) of the Josephson junctions written with an
	irradiation dose of $200\text{ ions/nm}$, which are measured at different
	temperatures of $40$, $50$, $60$ and $70\text{ K}$. At $40\text{ K}$, the critical current value is about $0.5\text{ mA}$, which corresponds to a critical current density $3.6\times10^5\text{ A/cm}^2$ according to the bridge 
	width and film thickness, while the normal resistance is about $0.8\ \Omega$, that results in the characteristic voltage value $V_{c} \simeq 0.4\text{ mV}$. 
	As shown in \cref{fig:HTS200Rn}, the normal
	resistance decreases with temperature decreasing and hence manifests a
	normal metal type of the junction barrier.
	
\begin{figure}[t!]
	\centering
	\subfloat[]{\includegraphics[width=7.2cm]{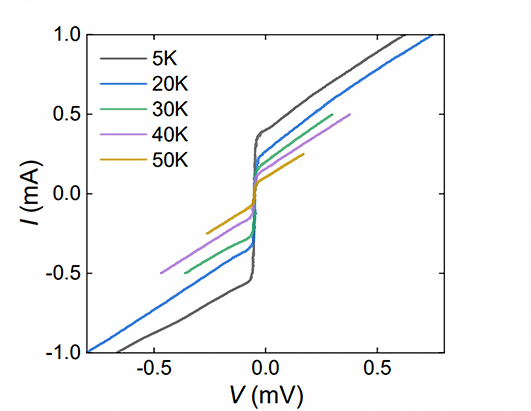}}\\
	\subfloat[]{\includegraphics[width=7.2cm]{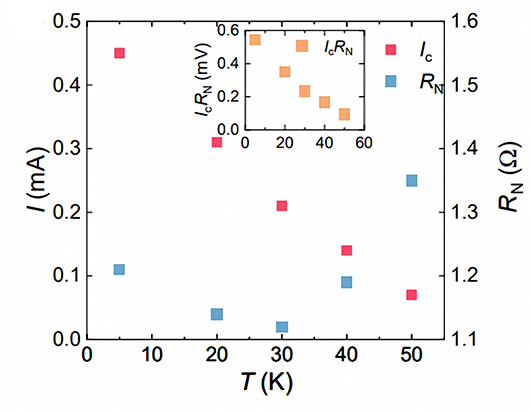}}
	\caption {\label{fig:HTS300Curves}IV curves (a), critical current $I_c$, normal
		resistance $R_N$ (b), and their product (characteristic voltage)
		$I_cR_N$ (inset) of the Josephson junctions fabricated using a helium dose of $300\text{ ions/nm}$, which are measured at different temperatures: $5$, $20$, $30$, $40$ and $50\text{ K}$.}
	\end{figure}

	\Cref{fig:HTS300Curves} shows IV curves (a), critical current $I_{c}$, normal
	resistance $R_{N}$, and their product (characteristic voltage)
	$I_{c}R_{N}$ (b) of the Josephson junctions written with a
	dose of $300\text{ ions/nm}$, which are measured at different temperatures from $5\text{ K}$ to $50\text{ K}$. At $40\text{ K}$, the critical current value is about $0.14\text{ mA}$, which corresponds to a critical current 
	density $1\times10^5\text{ A/cm}^2$, while the normal resistance is about $1.19\ \Omega$, that results in
	the characteristic voltage value $V_{c} \simeq 0.16\text{ mV}$. For this
	junction, the normal resistance first decreases but then increases with
	temperature decreasing, indicating a barrier property somehow between
	metal and insulator.
	
\begin{figure}[t!]
	\centering
	\subfloat[]{\includegraphics[width=7.2cm]{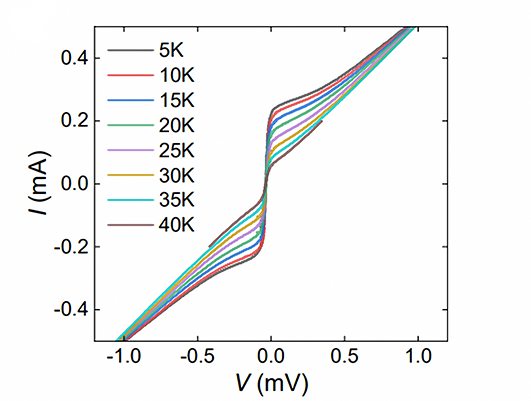}}\\
	\subfloat[]{\includegraphics[width=7.2cm]{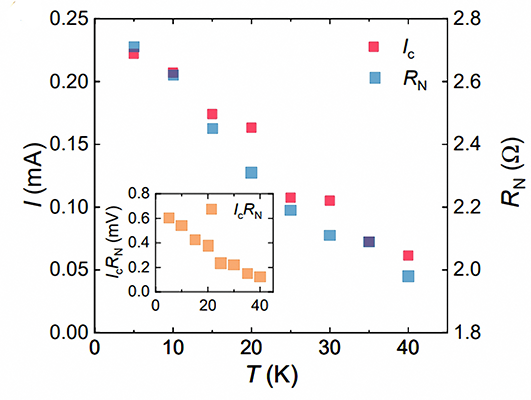}}
	\caption {\label{fig:HTS400Curves}IV curves (a), critical current $I_c$, normal
		resistance $R_N$ (b), and their product (characteristic voltage)
		$I_cR_N$ (inset) of the Josephson junctions fabricated using a helium dose of $400\text{ ions/nm}$, which are measured at different temperatures: $5$ to $40\text{ K}$.}
	\end{figure}

	Further increasing the helium dose, \cref{fig:HTS400Curves} shows IV curves (a),
	critical current $I_{c}$, normal resistance $R_{N}$, and their
	product (characteristic voltage) $I_{c}R_{N}$ (b) of the
	Josephson junctions fabricated using a dose of $400\text{ ions/nm}$, which are
	measured at different temperatures from $5$ to $40\text{ K}$. At $40\text{ K}$, the critical
	current value is about $60\ \mu\text{A}$, which corresponds to a critical current density $4.3\times10^4\text{ A/cm}^2$, while the normal resistance is
	about $2\ \Omega$, that results in the characteristic voltage value
	$V_{c} \simeq 0.12\text{ mV}$. For the junction written with a dose of 
	$400\text{ ions/nm}$, the normal resistance increases with temperature decreasing in
	contrast to the junctions written with a dose of $200\text{ ions/nm}$. This fact
	indicates rather insulator-close type of the junction barrier. Hence, by
	simply controlling the helium dose parameter, the barrier characteristic
	can be regulated gradually, which benefits the following multi-junction
	device designing.

	As one can see from \crefrange{fig:HTS200Curves}{fig:HTS400Curves}, 
	the fabricated HTS Josephson junctions
	show IV curves closed to the ones corresponding to the resistively
	shunted junction model (RSJ model) \cite{Likharev1986}of overdamped Josephson
	junctions (when the capacitive parameter, usually called as
	Stewart-McCumber parameter,
	\begin{equation}
		\label{eq:McCumberReq}
		\beta_{c} = \left({2e}/{\hslash} \right)I_{c}R_{N}^{2}C \ll 1,
	\end{equation}
	where $C$ is the junction capacitance) \cite{Likharev1986}. This is just the
	type of the junctions which are needed for designing a major part of the
	promising superconductor devises. Moreover, the obtained characteristic
	voltage values $V_{c} \simeq \left(0.12\text{ to }0.4\right)\text{ mV}$ 
	at $40\text{ K}$ are quite
	high and comparable with the ones of the resistively shunted Nb tunnel
	junctions. In fact, at the often-used critical current density $j_c=4.5\text{ kA/cm}^2$ and $I_c\simeq0.1\text {mA}$, requirement \cref{eq:McCumberReq} yields in $V_C\lesssim0.4\text{ mV}$ (see design parameters in \cite{SEEQC}). Besides, the reported values of $V_c$ are not so much less than the best 
	ones feasible for the grain-boundary junctions: $1$ to $3\text{ mV}$
	at temperature $4.2\text{ K}$ and $0.1$ to $0.3\text{ mV}$ at $77\text{ K}$ as summarized in \cite[p.~323]{Seidel2015}. One can expect more progress 
	in the Josephson junction parameters with further refinement of the fabrication technology.
	
	\section{Discussion}
	
	Main advantages of the focused helium ion beam fabrication technique
	developed and reported here are: (i) the achievable high value of the
	characteristic voltage $V_{c} = I_{c}R_{N}$ of the YBCO Josephson
	junctions; (ii) the controllable barrier characteristic; (iii) the
	ability to design a large number of arbitrary located Josephson
	junctions. These facts allow narrowing the existing gap in design
	abilities for LTS and HTS circuits even with using only single HTS film
	layer.

	\begin{figure}[b!]
		\centering
		\includegraphics[width=7.5cm]{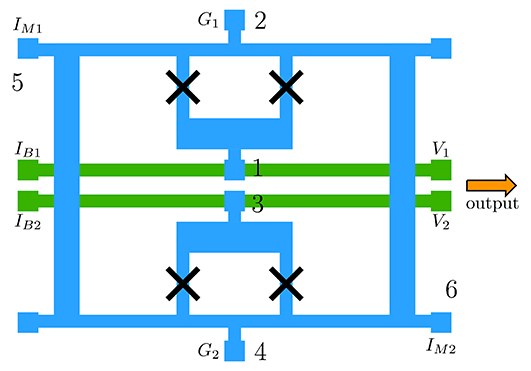}
		\caption{
			\label{fig:SingleLayer}Schematic layout of the one-layer active ESA based on using DSQUID
			\cite{Soloviev2019}. Superconductor film strips are shown by blue color, while
			normal metal lines (wires for connection) are shown by green color.
			The antenna contains quadratic superconducting strip loop and two dc SQUIDs inside the loop. The SQUIDs are biased independently by means 
			of two current sources, each applying dc bias current
			$I_{b} \simeq 2I_{c}$ are connected to points 1, 2 and 3, 4,
			respectively. The points 2 (G1) and 4 (G2) are to be connected via some resistors to a normal metal ground (this connection is not shown). Output signal is voltage $V$ measured between points 1, 3. An incident electromagnetic wave (namely, a flux of its
			magnetic field component) induces circular screening current in the
			superconducting loop. This current applies magnetic flux to both dc
			SQUIDs. An additional source of dc current $I_{m}$ is connected to
			points 5,6 to apply dc magnetic flux $\Phi_{m} \lesssim \Phi_{0}/4$ to
			each of the dc SQUIDs but with opposite signs. The antenna output
			voltage $V = V_{1} - V_{2}$ is measured between points 1,3.
		}
	\end{figure}
	
	Characteristic voltage $V_c$ of Josephson junctions is important figure of merit for device application. The increase in $V_c$	of the used Josephson junctions essentially improves performance of all Josephson junction devices, including single dc SQUIDs and more complicated SQUID-based devices, by increase \textit{(i)} in the output voltage and the flux-to-voltage transfer factor, both are proportional to $V_c$, and hence in dynamic range of the devices \cite{Kornev2017}, \textit{(ii)} in the maximum signal (averaged over Josephson oscillations) frequency $F_\text{sign} \ll F_c$ through increase in the characteristic Josephson-junction frequency $F_c$ \cref{eq:CharFreq}.
	
	One of such simple and highly promising devices which 
	can be realized using the single-layer topology is active 
	broadband receiving electrically small antenna (ESA) 
	based on using differential dc SQUID circuit (DSQUID) 
	\cite{Soloviev2019}. \Cref{fig:SingleLayer} shows a schematic layout of the active ESA 
	which can be formed on chip of very small size $a\times a$ such as $5$ by $5\text{ mm}^2$ or even less and therefore can be capable of receiving electromagnetic signals through sensing their magnetic component $B$ in the wide frequency band starting from dc.
	
	The antenna contains quadratic superconducting strip loop 
	and two dc SQUIDs inside the loop. The SQUIDs are biased independently by means of two current sources, each applying dc bias current $I_b=2I_c$, connected to points 1, 2 and 3, 4, respectively. The points 2 (G1) and 4 (G2) are  to be connected via some resistors to a normal metal 
	ground (this connection is not shown in \cref{fig:SingleLayer}). Output signal is voltage $V=V_1-V_2$ measured between points 1, 3. 
	
	A circular screening current is induced by an incident electromagnetic wave (namely, by a flux of its magnetic field component) in the superconducting loop, and therefore applies magnetic flux $\Phi_e$ to each of the dc SQUIDs. An additional source of dc current $I_m$ is connected to points 5, 6 
	to apply dc magnetic flux $\Phi_m\lesssim\Phi_0/4$ to each of the dc 
	SQUIDs but with opposite signs to shift voltage responses 
	of the dc SQUIDs in opposite directions.
	
\begin{figure}[t!]
		\centering
		\includegraphics[width=7cm]{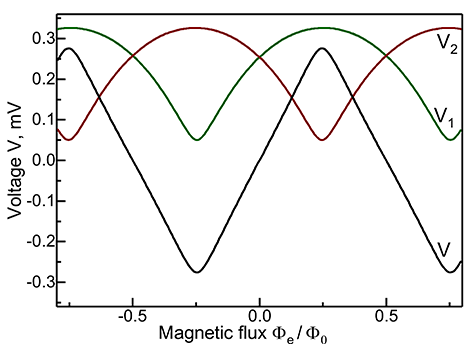}
		\caption{\label{fig:SingleLayerVR}
			Voltage responses $V_{1},\ V_{2}$ of both dc SQUIDs with $l = 2$
			and resulting voltage response $V = V_{1} - V_{2}$ of the antenna at
			$I_{b} = 2.1I_{c}$ obtained by means of numerical simulation using
			PSCAN software \cite{Polonsky1991,Polonsky1997}. Here $\Phi_{e}$ is the magnetic flux
			applied to each of dc SQUIDs; $\Phi_{e} = \Phi_\text{ex}M/L_\text{ant}$, where
			$\Phi_\text{ex}$ is the flux applied to the antenna loop area by an
			incident electromagnetic wave, $L_\text{ant}$ is the antenna loop
			inductance, and $M$ is the mutual inductance between the antenna
			loop and dc SQUID loop. Voltage scale is given for the case of using
			YBCO Josephson junctions with characteristic voltage
			$V_{c} = 0.4\text{ mV}$).}
	\end{figure}

	\Cref{fig:SingleLayerVR} shows the voltage responses $V_{1}$, $V_{2}$ of both dc SQUIDs (measured between points 1, 2 and 3, 4, respectively) and resulting voltage response of the antenna $V = V_{1} - V_{2}$ versus magnetic flux $\Phi_e$ applied to each of the 
	two dc SQUIDs. The responses were obtained by numerical simulation using PSCAN software \cite{Polonsky1991,Polonsky1997}.
	As far as form of the voltage response of dc SQUID at the critical 
	current biasing $I_b=2I_c$ is very close to parabolic law, the 
	difference of the mutually shifted voltage responses (by dc flux $\Phi_m$) ) voltage responses ($V_1$ and $V_2$), which 
	is the antenna voltage measured between points 1, 3, shows 
	linear response to the incident wave. The linear response 
	range allows the peak-to-peak value of the signal magnetic 
	flux $\Phi_e$ up to about $\Phi_0/2$.
	
	When normalized inductance of the used dc SQUIDs
	\begin{equation}
	l = \frac{2\pi I_{c}L}{\Phi_{0}}
	\end{equation}
	is much less than $1$, maximum swing of the antenna voltage 
	response $\Delta V$ can be at $I_b =2I_c$ as high as $2V_c$
	resulting in the flux-to-voltage transfer factor value
	\begin{equation}
		V_\Phi=dV/d\Phi_\text{ex}=4V_C/\Phi_0
	\end{equation}
	Both $\Delta V$ and $V_\Phi$ decrease with increase in the bias current $I_b$ and with increase in normalized inductance $l$ \cite{Likharev1986,Seidel2015}.
	
	Sensitivity of the antenna to the flux $\Phi_e$ applied to each 
	of the dc SQUIDs (internal sensitivity) is given by ratio of 
	the \textit{rms} value $V_f$ of the output voltage noise to the transfer 
	factor $V_\Phi$. In compliance with Langevin method, the noise 
	current component through Josephson junction (JJ) generated by its $R_N$ can be described through parallel connection of noiseless JJ and noise current source with the proper spectrum density $S_I$, which in the thermal noise limit is $S_I=4k_BT/R_N$, where $k_B$ is the Boltzmann’s constant 
	and $T$ is the ambient temperature. In view of independence 
	of the noise sources, the \textit{rms} value $V_f$ can be estimated,
	using approximation of the effective differential resistance 
	of JJ in resistive state by $R_N$, as follows:
	\begin{equation}
		V_f\simeq\left\{2\left[2S_IR_N^2/4\right]\Delta F\right\}^{1/2}=2\sqrt{k_BTR_N\Delta F},
	\end{equation}
	where $\Delta F$ is the frequency band of the subsequent registering unit. Thus, the internal sensitivity is as follows:
	\begin{equation}
		\Delta\Phi_e/\sqrt{\Delta F}=V_f/V_\Phi\approx\Phi_0\sqrt{k_BTR_N}/\left(2V_c\right).
	\end{equation}
	This characteristic improves with characteristic voltage $V_c$ of the used Josephson junctions.
	
	At a fixed size of dc SQUID loops and hence their inductance $L$, resulting sensitivity of the active ESA to the incident electromagnetic wave rises proportionally to the linear size a of the antenna loop, while the antenna remains electrically small (at $a \lesssim \lambda/6$). In fact, the signal magnetic flux $\Phi_e$ applied by the incident wave to each of the 
	dc SQUID s is proportional to the screening current induced in the antenna loop $I_\text{ant}=BS_\text{ant}/L_\text{ant}$, where $B$ is 
	the magnetic field component perpendicular to the antenna 
	chip plane, $S_\text{ant}$ is the antenna loop area, and 
	$L_\text{ant}$ is the antenna loop inductance. This current and hence 
	$\Phi_e$ both rise proportionally to the antenna size $a$ since 
	$S_\text{ant}$ is proportional to $a^2$, while 
	$L_\text{ant}$ is proportional to $a$.
	
	When $a\gtrsim\lambda/6$, one has to take into consideration an appeared inhomogeneity in the magnetic field distribution 
	along the antenna length. In case of a harmonic incident 
	wave, such a consideration results in decrease of the applied magnetic flux amplitude described by factor 
	\begin{equation}
		D=\sin\left(\pi a/\lambda\right)/\left(\pi a/\lambda\right).
	\end{equation}
	This factor gives a $3\text{ dB}$ reduction in the flux amplitude and 
	hence in amplitude of the antenna output voltage at
	$\lambda_0\simeq9a/4=2.25a$. Thus, the value $\lambda_0$ sets
	the upper frequency of the antenna band starting from dc
	
	Further development of the active electrically small antenna (ESA) can
	be obtained by replacing two dc SQUIDs by two parallel arrays of
	Josephson junctions, i.e. with realizing a particular case of active
	ESA based on SQA (superconducting quantum array) \cite{Kornev2014} composed of
	differential unit blocks \cite{Kornev2017}. Such a design approach can be used
	also to develop high-performance rf amplifiers.
	
	The use of multi-junction array structures like SQAs \cite{Kornev2014} or SQIF
	\cite{Caputo2005,Oppenlander2001} enables an increase of 
	overall device performance, including
	increase in dynamic range of the multi-junction devices, through
	increase in signal-to-noise ratio (SNR) with the number $N$ of unit
	cells proportional to $\sqrt{N}$. In fact, as long as intrinsic
	fluctuations in array cells are independent, spectral density of the
	low-frequency voltage fluctuations (at the signal frequency $\Omega$)
	across the serially connected $N$ cells
	$S_{v}(\Omega) = NS_{V}^{0}(\Omega)$, where $S_{V}^{0}(\Omega)$ is
	the spectral density across one cell. Thus, the \emph{rms} fluctuation
	voltage $V_{F} = \sqrt{NS_{V}^{0}(\Omega)\Delta\Omega}$ , where
	$\Delta\Omega$ is the frequency band, increases as $\sqrt{N}$,
	whereas both the maximum output signal $V_{\max}$ and transfer factor
	$V_{\Phi} = dV/d\Phi$ of the flux $\Phi$ to voltage $V$ transformation
	increases as $N$. Therefore, dynamic range $DR = V_{\max}/V_{F}$
	increases as $\sqrt{N}$. In case of $N$ cells connected in
	parallel, both the maximum output voltage $V_\text{max}$
	and transfer factor $V_\Phi$ don't change with
	$N$, while the spectral density of the low frequency current
	fluctuations becomes $N$ times higher than the one for one cell
	$S_{I}^{0}(\Omega)$, and therefore the \emph{rms} fluctuation voltage
	$V_{F} = R_{d}^{0}\sqrt{NS_{I}^{0}(\Omega)\Delta\Omega}/N$,
	where $R_{d}^{0} = dV/dI$ is differential resistance at operation
	point on IV curve of one cell, decreases as $1/\sqrt{N}$, and hence
	dynamic range $DR = V_{\max}/V_{F}$ increases also as $\sqrt{N}$.
	
	\section{Conclusion}
	
	The obtained progress in the HTS (YBCO) Josephson junction fabrication
	process founded on using a focused helium ion beam damaging technique
	enables achieving characteristic voltage value $V_{c} = I_{c}R_{N}$
	up to about $0.4\text{ mV}$ and potentially higher. Both the high value of
	$V_{c}$ of Josephson junctions and the ability to design a large
	number of arbitrary located Josephson junctions allow narrowing the
	existing gap in design abilities for LTS and HTS circuits even with
	using a single HTS film layer. A one-layer topology of active
	electrically small antenna is suggested and the antenna voltage response
	characteristics are considered. Such a design approach can be also used
	to develop high-performance rf amplifiers. Further development of the
	active devices can be obtained by using multi-junction array structures
	like SQAs which enables an increase in overall performance, including
	increase in dynamic range of the multi-junction devices, through
	increase of signal-to-noise ratio (SNR) with the number $N$ of unit
	cells proportional to $\sqrt{N}$.

	\bibliographystyle{emsreport}
	\bibliography{HTS_MultiJunctions}
	\end{document}